\documentclass[a4paper,11pt]{article}
\pdfoutput=1 % if your are submitting a pdflatex (i.e. if you have
\usepackage{jheppub} % for details on the use of the package, please
\usepackage[T1]{fontenc} % if needed
\usepackage{amsmath}
\usepackage{amsfonts}
\usepackage{amssymb}
\usepackage{graphicx}
\usepackage{hyperref}
\usepackage{color}
\usepackage{caption}
\usepackage{verbatim}
\usepackage[normalem]{ulem}
\usepackage{subfigure}
\usepackage{slashed}
\usepackage{cancel}
\usepackage{braket}
\usepackage{multicol}
\usepackage{tcolorbox}
\tcbuselibrary{theorems}
\usepackage{bm}
\usepackage{breqn}

\usepackage{physics}
\usepackage{tikz-feynman}

\def\micro{{\tt micrOMEGAs}}

\def\darkON{{\tt darkOmegaN}}
\def\darkO{{\tt darkOmega}}
\def\exclu{{\tt Excluding2010}}
\def\mad{{\tt Madgraph5 v1.5.8}}

\title{\boldmath Thermal Dark Photon Dark Matter, Coscattering, and Long-lived ALPs}
\author[a]{Bastián Díaz Sáez,}
\affiliation[a]{Instituto de Física, Pontificia Universidad Católica de Chile,
Avenida Vicuña Mackenna 4860, Santiago, Chile}
\emailAdd{bastian.diaz@uc.cl}
\abstract{We study the thermal freeze-out of a dark photon dark matter in the so-called \textit{dark-axion portal} - a triple coupling among a dark photon, an axion-like particle and the SM $\gamma$ or $Z$ boson. We analyze in detail the thermal production regimes: coscattering (aka conversion driven freeze-out), mediator freeze-out, and coannihilations. We found viable DM scenarios fulfilling the correct relic abundance in the three regimes. Apart of analyzing the parameter space for dark state masses between the GeV-TeV scale, we explore the prospects of having the axion-like field as a long-lived particle, possibly to be observed at the LHC and future detectors.}

\begin{document} 
\maketitle
\flushbottom

\section{Introduction}
Many extensions to the Standard Model of particle physics predict the existence of new bosons of spin 0 and 1. The interplay of dark photons (DP) and axion-like particles (ALP) in the so-called \textit{dark-axion portal}, this is, when the ALP couples with a DP and a SM photon, has been the subject of several studies, ranging from dark matter \cite{Kaneta:2016wvf, Kaneta:2017wfh}, astrophysical probes \cite{Hook:2021ous}, cosmology \cite{Arias:2020tzl, Gutierrez:2021gol, Hong:2023fcy, Hook:2023smg}, and collider experiments \cite{deNiverville:2018hrc, deNiverville:2019xsx, Deniverville:2020rbv, Jodlowski:2023sbi}. In particular, a few works have situated this framework in the freeze-in regime, for instance, with the DM production coming from either QCD or dark Primakov \cite{Kaneta:2017wfh}, with sub-GeV masses for the dark states, with both dark particles behaving as feebly states \cite{Arias:2025nub}, and via the decay of a thermalized ALP \cite{Arias:2025abd}. 

In this work, and for the very first time, we study the freeze-out of a dark photon as a dark matter (DM) in the dark-axion portal in the thermal regime for GeV-TeV mass scales. That is, we assume that the DP and the ALP were in thermal equilibrium at high temperatures with the SM plasma until both particles freeze-out. In particular, we pay special attention to coscattering \cite{DAgnolo:2017dbv}, also known as conversion-driven freeze-out \cite{ Garny:2017rxs} (for other scenarios based on coscattering, see \cite{Cheng:2018vaj, Brummer:2019inq, Junius:2019dci, Heeck:2022rep, Heisig:2024xbh, DiazSaez:2024nrq}). Coscattering relies on the fact that the DM particle and a heavier dark state are in chemical equilibrium at high temperatures via fast collisions with relativistic SM particles in the thermal bath. That is, the chemical potentials within the dark sector are the same, but once the reaction rate of these interactions drops below the Hubble expansion rate, the chemical equilibrium within the dark sector no longer supports, and the DM freezes out.

In order to solve the coupled Boltzmann equations (cBE) for the yields of dark sector particles, we use the \micro\ code \cite{Belanger:2001fz}. In it, some specific functions solve the cBE numerically. In particular, the new function \darkON\ calculates the relic abundance of the DM candidate without assuming CE during the whole decoupling process, unlike the old \micro\ function \darkO\, which is also used in this work to compare the results. 

In some regions of the parameter space, especially when coscattering is the dominant thermal regime, the ALP lifetime significantly increases to the point to be considered as a \textit{long-lived} particle (LLP). This makes coscattering a mechanism that is phenomenologically testable at colliders \cite{Curtin:2018mvb, Cottin:2024dlo}. In this way, we scan some regions of the parameter space, founding that, in principle, the physics of the dark-axion portal could be tested via LLP at present and future detectors via, for instance, non-pointing photons.

The paper is structured in the following way. In Sect.~\ref{sec1}, we present the model. In Sect.~\ref{sec_relic}, we study in detail the relic abundance of this scenario. In Sect.~\ref{collider_pheno}, we present the collider phenomenology of this scenario, focusing on the ALP as a LLP. Finally, in Sect.~\ref{sec4} we present the conclusions.

\section{Model}\label{sec1}
We extend the gauge sector of the SM with a $U(1)'$ gauge symmetry. As a result, there is a new gauge field $A_\mu'$, and also we introduce a real pseudoscalar $\phi$. We set that this scenario has a charge conjugation symmetry (also called dark CP symmetry) acting on the new fields as $\phi \rightarrow -\phi$ and $A'_\mu \rightarrow - A'_\mu$, and the SM fields transforming trivially under it \cite{Hook:2021ous} (also see \cite{Ma:2017ucp, Duerr:2018mbd}). In this way, the Lagrangian of this SM extension is given by:
\begin{eqnarray}\label{lag1}
 \mathcal{L} \supset - \frac{1}{4}B_{\mu\nu}B^{\mu\nu} - \frac{1}{4}F'_{\mu\nu}F^{'\mu\nu} - \frac{1}{2}m_{\gamma'}^2A_\mu^{'2} + \frac{1}{2}(\partial_\mu \phi)^2 - \frac{1}{2}m_\phi^2\phi^2 - \lambda_\phi \phi^4 - \lambda_{HS}\phi^2 |H|^2,
\end{eqnarray}
with $(m_\phi, \lambda_{\phi}) > 0$, $B_{\mu\nu}$ the $U(1)_Y$ photon field strength, and $H$ the Higgs doublet of the SM. Due to the symmetry of this scenario, there is no kinetic mixing between $B_\mu$ and $A'_\mu$, and it is allowed the following five-dimensional operator\footnote{In the first version of this work all the calculation was using the $U(1)_{em}$ strength field. Even when the changes in the present version are not too strong, the presence of the $Z$ boson is necessary for the correct calculation once the dark state masses are above the GeV scale.}: 
\begin{eqnarray}
\mathcal L_{5}= \frac{g_D}2 \phi F'_{\mu\nu}\tilde B^{\mu\nu},
\label{eq:portal}
\end{eqnarray}
where $g_D$ is a dimensionful parameter related to a high energy scale as $f_D = 1/g_D$, and $\tilde B^{\mu\nu}$ being the dual field strength defined as $\tilde B^{\mu\nu} = \frac{1}{2}\epsilon^{\mu\nu\alpha\beta}
B_{\alpha\beta}$. The new singlet scalar $\phi$ behaves as an unstable state which does not acquire any vacuum expectation value.

In the following, we recognize $\phi$ as an axion-like particle (ALP), that is, a pseudo-Nambu Goldstone boson whose shift symmetry has been broken by UV physics. For example, this can be realized in a hidden $SU(3)$ gauge model, where below its confinement scale, new hidden fermions condense forming the ALP $\phi$. Furthermore, if it is allowed a Yukawa-like term between the Higgs doublet and hidden $SU(3)$ fermions, the resulting low energy description could give rise to a Higgs portal term as the last one in eq.~\ref{lag1}, with the allowed values of $\lambda_{HS}$ model dependent (e.g.  \cite{Jeong:2018ucz, Im:2019iwd}). Realistic UV completion is beyond the scope of this work.

The hierarchy $m_\phi < m_{\gamma'}$ makes the ALP absolute stable. In this case, in order to cover all the thermal phases for the ALP, one must vary the two couplings $\lambda_{HS}$ and $g_D$, since both are responsible for the interaction between the ALP and the SM sector. Although the study of the latter case is possible, it is more involved than the case $m_{\gamma'} < m_\phi$, where the dark photon becomes the candidate for DM. In the latter case, there is a single coupling connecting the DM and the visible sector, $g_D$, while $\lambda_{HS}$ behaves as a coupling regulating only the interaction of the mediator $\phi$ with the SM sector. 

Last but not least, we analyze the $m_{\gamma'} < m_\phi$ hierarchy not only by its simplicity over the other hierarchy, but because $\phi$ may behave as a LLP at colliders, and sizable values of $\lambda_{HS}$ have two purposes: i) to pair produce the ALPs with sizable cross sections at colliders, and ii) to keep $\phi$ in thermal equilibrium in the early universe. In the rest of the paper we consider independent parameters $(m_{\gamma'},m_\phi, g_D, \lambda_{HS})$, and we make use of the parameter $\Delta m \equiv m_\phi - m_{\gamma'}$.

\section{Relic abundance}\label{sec_relic}
In this section, we solve the set of coupled Boltzmann equations (cBE) of $\gamma'$ and $\phi$ using \micro \ 6.0.4, which assumes that both particles were in CE at high temperatures. We implement this scenario by assuming three thermal systems: the SM, the DM candidate $\gamma'$, and $\phi$ the heavier state.

\subsection{Boltzmann equations}
Introducing the independent variable $x = m_{\gamma'}/T$, the cBE for the yields of $\gamma'$ and $\phi$, $Y_{\gamma'}
$ and $Y_\phi$, respectively, are given by
%+ \ev{\sigma_{\gamma'\gamma'\phi\phi} v}\left(Y_{\gamma'}^2 - Y_\phi^2\frac{Y_{\phi ,e}^2}{Y_{\gamma' e}^2}\right)
\begin{dmath}\label{beq1}
 \frac{dY_{\gamma'}}{dx}  = \frac{1}{3H}\frac{ds}{dx}\left[\ev{\sigma_{\gamma'\gamma' 00} v}\left(Y_{\gamma'}^2 - Y_{\gamma',e}^2\right) + \ev{\sigma_{\gamma'\phi 00} v}\left(Y_{\gamma'}Y_\phi - Y_{\gamma' e}Y_{\phi, e}\right) \\   + \frac{\Gamma_{\gamma'\rightarrow \phi}}{s}\left(Y_{\gamma'} - Y_\phi\frac{Y_{\gamma'e}}{Y_{\phi e}}\right) + \frac{\Gamma_\phi}{s}\left(Y_\phi - Y_{\gamma'}\frac{Y_{\phi e}}{Y_{\gamma' e}}\right)\right],
 \end{dmath}
 
 \begin{dmath} \label{beq2}
  \frac{dY_\phi}{dx} = \frac{1}{3H}\frac{ds}{dx}\left[\ev{\sigma_{\phi\phi 00} v}\left(Y_\phi^2 - Y_{\phi, e}^2\right) + \ev{\sigma_{\gamma' \phi 00} v}\left(Y_{\gamma'}Y_\phi - Y_{\gamma' e}Y_{\phi , e}\right) \\ - \frac{\Gamma_{\gamma'\rightarrow \phi}}{s}\left(Y_{\gamma'} - Y_\phi\frac{Y_{\gamma' e}}{Y_{\phi, e}}\right)-\frac{\Gamma_\phi}{s}\left(Y_\phi - Y_{\gamma'}\frac{Y_{\phi, e}}{Y_{\gamma' e}}\right)\right],
\end{dmath}
where $0$ denotes SM particles, and the equilibrium yields are
\begin{eqnarray}
    Y_{\gamma',e}(x) &=& \frac{45}{4\pi^4}\frac{x^2}{g_{*S}(x)}K_2(x), \\
    Y_{\phi ,e}(x) &=& \frac{45}{4\pi^4}\frac{1}{g_{*S}(x)}\left(\frac{m_\phi x}{m_\gamma'}\right)^2K_2\left(\frac{m_\phi x}{m_\gamma'}\right),
\end{eqnarray}
\begin{figure}
    \centering
    \begin{tikzpicture}[baseline=-.1em, scale=.45]
        \begin{feynman}[small]
            \vertex (S2) at (-2,2) {$\gamma'$};
            \vertex (S1) at (4, 2) {$\phi$};
            \vertex (f1) at (-2,-2) {$\gamma, Z$};
            \vertex (f2) at (4,-2) {$h$};
            \vertex (a) at (0, 0);
            \vertex (b) at (2, 0);
            \diagram* {
                (S1) -- [scalar] (b) -- , (a) -- [boson] (S2),
                (a) -- [scalar, edge label'=$\phi$, momentum=$p_1 + p_2$] (b),
                (f1) -- [boson] (a) -- ,(b) -- [scalar] (f2)
            };
        \end{feynman}
    \end{tikzpicture}\hspace{1cm}
    \begin{tikzpicture}[baseline=-.1em, scale=.45]
        \begin{feynman}[small]
            \vertex (S2) at (-2,2) {$\gamma'$};
            \vertex (S1) at (2, 2) {$\phi$};
            \vertex (f1) at (-2,-2) {$f$};
            \vertex (f2) at (2,-2) {$f$};
            \vertex (a) at (0, 1);
            \vertex (b) at (0,-1);
            \diagram* {
                (S1) -- [scalar] (a) -- [boson] (S2),
                (a) -- [boson, edge label'=$\gamma$ $Z$, momentum=$p_1-p_2$] (b),
                (f1) -- [fermion] (b) -- [fermion] (f2)
            };
        \end{feynman}
    \end{tikzpicture}\hspace{0.5cm}
    \begin{tikzpicture}[baseline=-.1em, scale=.45]
        \begin{feynman}[small]
            \vertex (S2) at (-2,2) {$\gamma'$};
            \vertex (S1) at (2, 2) {$\phi$};
            \vertex (f1) at (-2,-2) {$W^\pm$};
            \vertex (f2) at (2,-2) {$W^\pm$};
            \vertex (a) at (0, 1);
            \vertex (b) at (0,-1);
            \diagram* {
                (S1) -- [scalar] (a) -- [boson] (S2),
                (a) -- [boson, edge label'=$\gamma$ $Z$, momentum=$p_1-p_2$] (b),
                (f1) -- [boson] (b) -- [boson] (f2)
            };
        \end{feynman}
    \end{tikzpicture}\hspace{0.5cm}
    \begin{tikzpicture}[baseline=-.1em, scale=.45]
        \begin{feynman}[small]
            \vertex (S2) at (-2,2) {$\gamma'$};
            \vertex (S1) at (2, 2) {$\gamma, Z$};
            \vertex (f1) at (-2,-2) {$h$};
            \vertex (f2) at (2,-2) {$\phi$};
            \vertex (a) at (0, 1);
            \vertex (b) at (0,-1);
            \diagram* {
                (S1) -- [boson] (a) -- [boson] (S2),
                (a) -- [scalar, edge label'=$\phi$, momentum=$p_1-p_2$] (b),
                (f1) -- [scalar] (b) -- [scalar] (f2)
            };
        \end{feynman}
    \end{tikzpicture}
    \caption{Tree-level contributions to $\gamma'\to \phi$ conversions.}
    \label{fig:conversion-diagrams}
\end{figure}
with $K_1(x)$ and $K_2(x)$ the Bessel second order functions, $g_{*S}(x)$ the effective number degrees in entropy, $s = \frac{2\pi^2}{45} g_{*S}T^3$ the entropy density in the universe, and $H$ the Hubble rate in a radiation-dominated universe. The rate of particle conversions per DM particle is given by
\begin{eqnarray} \Gamma_{\gamma'\rightarrow \phi} = \sum_{k,l}\ev{\sigma_{\gamma' k\rightarrow \phi l} v} n_{k,e},
\end{eqnarray}
with $k$ and $l$ being a SM state (see the diagrams in Fig.~\ref{fig:conversion-diagrams}), and $\Gamma_\phi$ is the thermally averaged decay rate of $\phi$. As the SM photon is massless, all the $\phi$ decays in this scenario are on-shell. 

Finally, we consider the correct relic abundance as the one measured by Planck collaboration $\Omega_c h^2 = 0.12$ \cite{Planck:2018vyg}, which theoretically is obtained by 
\begin{eqnarray}
    \Omega h^2 = \frac{2.9713\times 10^9}{10.5115\text{ GeV}}m_{\gamma'} Y_{\gamma',\infty},
\end{eqnarray}
with $Y_{\gamma',\infty}$ the yield obtained after the freeze-out of the DM.

%$\Delta_{2s}$ quantifies the degree of relevance of processes $10\leftrightarrow 20$ in the calculation of relic abundance, considering decays. 

\subsection{Relic abundance}
\begin{figure}[t!]
\centering
\includegraphics[width=0.45\textwidth]{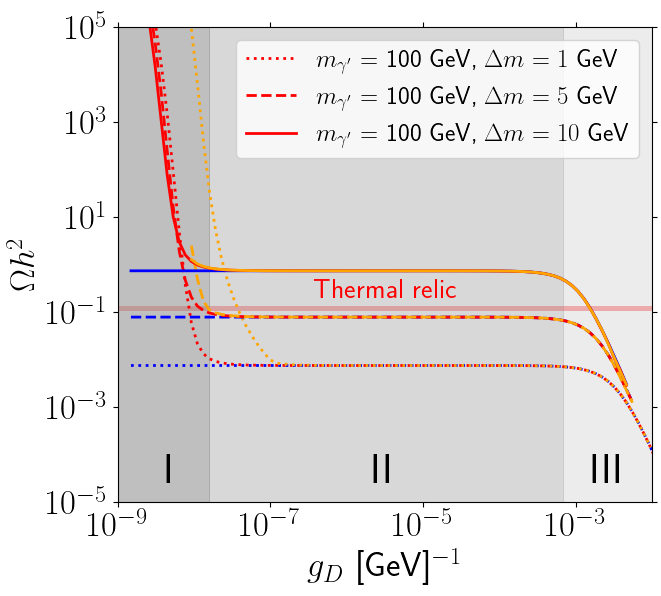}
\caption{Relic abundance as a function of $g_D$ for $m_{\gamma'} = 100$ GeV and different mass shift. The red curves are obtained with \darkON , the blue ones with \darkO, and the orange ones without considering processes 1020 in eqs.~\ref{beq1} and \ref{beq2}. The regions I, II and III, correspond to the particular case $\Delta m = 5$ GeV. We have set $\lambda_{HS} = 1$.} 
\label{plot1}
\end{figure}

As $T\lesssim (m_{\gamma'}, m_\phi)$, and if the coupling $g_D$ is sufficiently small, the CE between $\gamma'$ and $\phi$ is broken, so that one must solve the cBE shown in Eqs.~\ref{beq1} and \ref{beq2}. In that regime, the heavier state $\phi$ remains in CE with SM longer, since it presents additional interactions via the Higgs portal. For simplicity, we fix $\lambda_{HS} = 1$, although later on we relax this assumption. In this way, the relic abundance depends only on three parameters $(m_{\gamma'},m_\phi, g_D)$. As we focus on the EW mass scale for the new states, we scan the relic abundance for $m_{\gamma'} = 100$ GeV as a function of $g_D$ and $\Delta m$, with the results shown in Fig.~\ref{plot1}

The red curves indicate the relic abundance obtained with \darkON . Three regimes are represented by the gray regions, typically known as coscattering (region I), mediator freeze-out (region II), and DM (co)annihilations (region III). In the coscattering regime, the relic abundance in each case depends strongly on $g_D$ and too slightly on the mass difference, independent of the mass of the DM. For comparison, for each $\Delta m$ we also show the results obtained with \darkO\ (blue lines), where the underlying assumption of this solver is CE in the full parameter space. In this way, when the blue and red curves match, CE
is always maintained: regions II and III. For example, in region II, the relic abundance is determined mainly by the annihilation of a pair of $\phi$, with $\ev{\sigma_{eff} v} \propto e^{-2x_f\Delta m/m_{\gamma'}} \ev{\sigma_{\phi\phi 00} v}$, with $x_f \approx m_{\gamma'}/T_f \approx 25$. In this way, it indicates that the larger the masses of the dark sector, the smaller the spread of the relic abundances for a fixed $\Delta m$. Even when in Fig.~\ref{plot1} we show the results of a particular case, $m_{\gamma'} = $ 100 GeV, we have checked that $\Delta m\lesssim 10$ GeV is required in the full parameter space, otherwise an overabundance is obtained. 

Regarding the importance of decays when exist CE between the DP and the ALP, we have also added the results obtained with \darkON \ without conversion processes $\gamma'0 \leftrightarrow \phi 0$ by invoking the \micro\ function \exclu. The results are shown by the orange lines in Fig.~\ref{plot1}\footnote{In certain region of too small $g_D$, the orange curves cut-off, since \micro\ is not able to find an initial temperature where the dark and SM sectors were in thermal equilibrium.}. To maintain CE within the dark sector in the absence of these types of scattering processes, decays keep CE within the dark sector, but require larger couplings $g_D$, and the effectiveness of decays to maintain CE increases when increasing $\Delta m$ due to the increase of the decay phase space.

\begin{figure}[t!]
\centering
\includegraphics[width=0.8\textwidth]{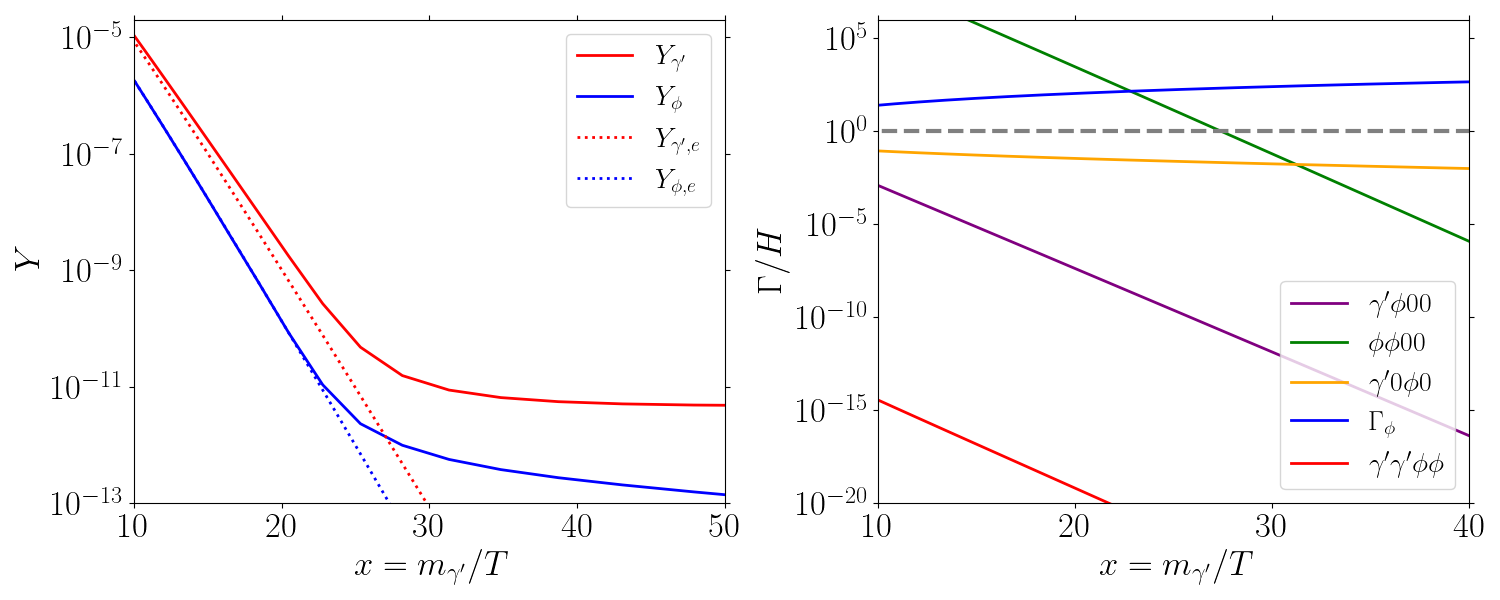}
\caption{(left) Yield evolution for the DP and the ALP as a function of the inverse of the temperature. Here we consider $m_{\gamma'} = 100$ GeV, $\Delta m = 5$ GeV, $g_D = 10^{-8}$ GeV$^{-1}$ and $\lambda_{HS} = 1$. (right) Particle interaction rates over Hubble rate as a function of the inverse temperature. "0" here refers to a SM particle.} 
\label{plot2}
\end{figure}

As in this work, we pay more attention to the coscattering regime, in Fig.~\ref{plot2}~(left) we show the yield evolution for a point in region I of Fig.~\ref{plot1}: $(m_{\gamma'}, m_{\phi}) = (100, 105)$ GeV and $g_D = 10^{-8}$ GeV$^{-1}$. Notice the early yield departure of the DP from the equilibrium $Y_{\gamma',e}$, typical of the coscattering regime, whereas the departures of $Y_\phi$ from its equilibrium occur later $x$, since $\phi$ is in CE via the Higgs coupling. In Fig.~\ref{plot2}~(right), we show the ratio between the interaction rates for the relevant processes over the Hubble expansion rate. As noted, conversions (orange line) remain below the Hubble rate for all $x$, thereby not guaranteeing CE within the dark sector.

\begin{figure}[t!]
\centering
\includegraphics[width=1.0\textwidth]{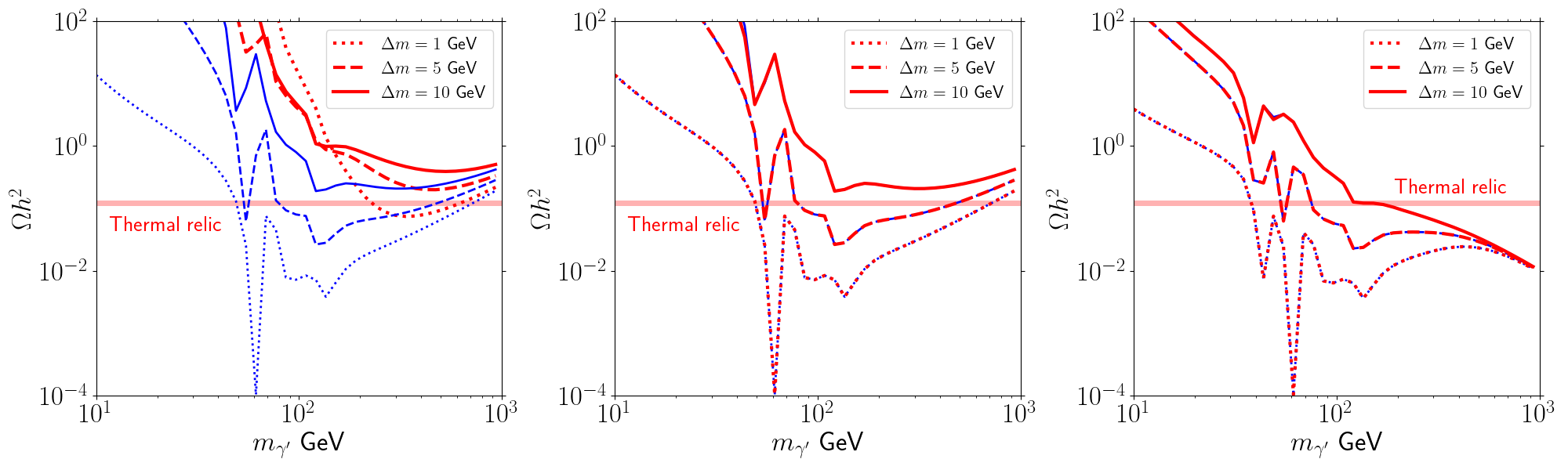}
\includegraphics[width=1.0\textwidth]{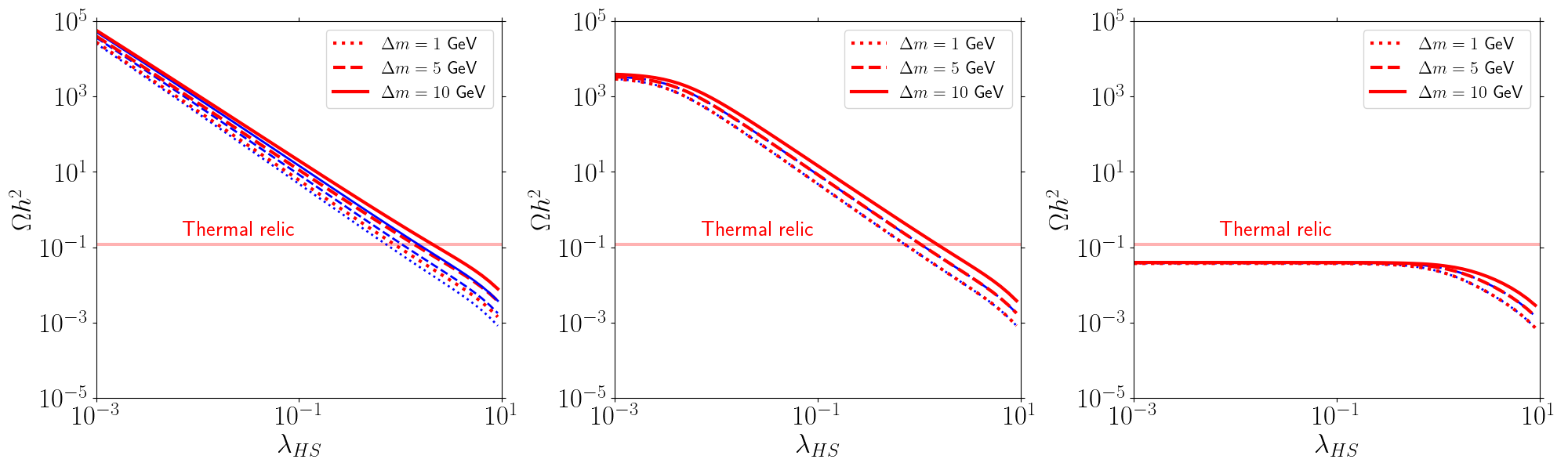}
\caption{(top) Relic abundance as a function of $m_{\gamma'}$. In the left, middle and right columns we consider $g_D = 6\times 10^{-9}$ GeV$^{-1}$, $10^{-5}$ GeV$^{-1}$ and $10^{-3}$ GeV$^{-1}$, respectively. The red and blue lines are obtained with \darkON \ and \darkO, respectively. Here we set $\lambda_{HS} = 1$. (bottom) Relic abundance as a function of $\lambda_{HS}$, for $m_{\gamma'} = 500$ GeV. The values of $g_D$ follows the same order than the top row of plots.} 
\label{plot3}
\end{figure}

To obtain a much more broad picture of the behavior of relic abundance as a function of the parameters, in Fig.~\ref{plot3} we show the dependence of relic abundance on the mass of the DP (top row) and on $\lambda_{HS}$ (bottom row), with the plot columns having different values of $g_D$. Notice that the chosen values for $g_D$ in each column were based on the typical values found in the first scan (see Fig.~\ref{plot1}) in the corresponding regions I, II and III, respectively. In the top left plot, as $g_D = 6\times 10^{-9}$ GeV$^{-1}$, there are sharp differences obtained with \darkON\ (red lines) and \darkO\ (blue lines), where the former gives the correct result, since CE is lost in this regime. In the middle and right plots of the first row, the match between the red and blue curve is practically in the whole range of masses, due to the fact that CE is recovered for such values of $g_D$. The inverted peak at $m_h/2$ in the second and third plot becomes sharper as $\Delta m$ decreases, since in that regime coannihilations are stronger, and therefore the annihilation $\phi\phi$ via an on-shell Higgs becomes efficient. For higher values of DP masses, $m_{\gamma'} \gtrsim 100$ GeV, the overall resulting effect of increasing $g_D$ varies depending on the thermal regime. For instance, as (co)annihilations are not effective in the mediator FO regime (middle plot), increasing $m_{\gamma'}$ makes that the relic abundance tend to increment independent on $\Delta m$, contrary to the decreasing behavior of the relic in the plot in the right, where (co)annihilations become more effective.

On the other hand, in Fig.~\ref{plot3}~(bottom), we observe deviations in the relic abundance as a function of $\lambda_{HS}$ for a fixed $m_{\gamma'} = 500$ GeV. The constant relic abundance is simply due to the fact that there are other processes participating in the freeze-out that do not depend on $\lambda_{HS}$, therefore dominating the relic abundance calculation for small enough values of this parameter. % Esta explicación era para el caso $g = 3E-9$, pero finalmente opté por poner el caso 6E-9In the bottom left plot, for high values of $\lambda_{HS}$, the differences between the results between \darkO\ and \darkON\ are due to the fact that, in the former case, annihilation FO increases monotonically, in particular via the leading process $\phi\phi\rightarrow W^+W^-$, whereas in the latter case, the processes depending on $\lambda_{HS}$ are subleading, e.g. $\phi h \rightarrow \gamma' \gamma$, then the changes with $\lambda_{HS}$ are mild.
From these three graphs, we observe that only certain sizable values for $\lambda_{HS}$ give the correct relic abundance, and in certain cases, as shown by the last graph, none of the values of $\lambda_{HS}$ will give the correct relic abundance. We do not expect strong deviations of these conclusions if we change $m_{\gamma'}$ in the mass range between $m_h/2$ and 1 TeV, with the exception of the Higgs resonance, where more sharp numerical deviations are present.

\section{Phenomenology}\label{collider_pheno}
In this section, we study general prospects to test the thermal dark-axion portal at present and future collider detectors. In the following, we discuss the proper lifetime of ALPs inspired by ATLAS at the LHC \cite{Lee:2018pag}, and we highlight a few possibilities to test signals from the pair production of ALPs as long-lived particles (LLP) via the Higgs portal eventually produced at the LHC. Finally, we comment on possible signal identification at the future MATHUSLA detector MATHUSLA.

Before continuing, here are just a few words about direct and indirect detection of the dark-axion portal. As we focus on coscattering and mediator FO, and typically those regimes present $g_D \ll 10^{-3}$ GeV$^{-1}$, the leading one-loop cross section relevant for direct detection is expected to be highly suppressed, thereby not imposing a relevant constraint in the parameter space studied here\footnote{In the opposite hierarchy, where the ALP is the lightest stable state  $m_\phi < m_{\gamma'}$, then the DM candidate, $\phi$ could be subject to strong tree-level direct detection rates through the Higgs portal. This case is not addressed in this work.}. On the other hand, it is possible to have thermal DM in the sub-GeV mass range considering $\Delta m < 1$ GeV. Preliminary results indicate that it is possible to obtain thermal dark photons with masses in the sub-GeV range for $\Delta m \approx 1$ MeV. In particular, this region of the parameter space could also be interesting in the context of direct detection with \textit{Inelastic DM} \cite{Tucker-Smith:2001myb}. Finally, with respect to indirect detection, SM fluxes born from the annihilation of a pair of DP occur at one loop level, and considering further suppression from $g_D$, we expect no relevant bounds from this type of searches in the coscattering nor in the mediator FO regimes, unless some non-perturbative effect enhanced the annihilation cross section.

\subsection{Long-lived particles}
Coscattering and mediator FO regimes present $g_D$ and $\Delta m$ small enough to establish the ALP $\phi$ as a long-lived particle (LLP). In our framework, the decay width of the ALP is given by
\begin{eqnarray}\label{decay_width}
    \Gamma_\phi = \frac{g_D^2c_W^2}{32\pi}m_\phi^3\left(1 - \frac{m_{\gamma'}^2}{m_\phi^2}\right)^3,
\end{eqnarray}
where $c_W$ is the cosine of the weak mixing angle, and then the proper decay length of $\phi$ given by
\begin{eqnarray}
    c\,\tau_{\phi}^0 = 1.97\times 10^{-14}\, \text{cm} \left(\frac{\text{GeV}}{\Gamma_\phi}\right)~.
\end{eqnarray}
For typical values of $\Gamma_\phi$ in the coscattering and mediator FO, $c\tau_\phi^0$ can take values such that a pair of $\phi$ produced at the LHC can decay inside the electromagnetic calorimeter (ECAL), the hadronic calorimeter (HCAL), and the muon spectrometer of the ATLAS experiment. To quantify this more precisely, we scan $m_{\gamma'} = [m_h/2, 500]$ GeV and $\Delta m = [1,10]$ GeV, maintain $\lambda_{HS} = 1$, and take a few values of $g_D$ in the range $[6\times 10^{-9}, 3\times 10^{-8}]$ GeV$^{-1}$. We select all the points which fulfill the correct relic abundance.

\begin{figure}[t!]
\centering
\includegraphics[width=1\textwidth]{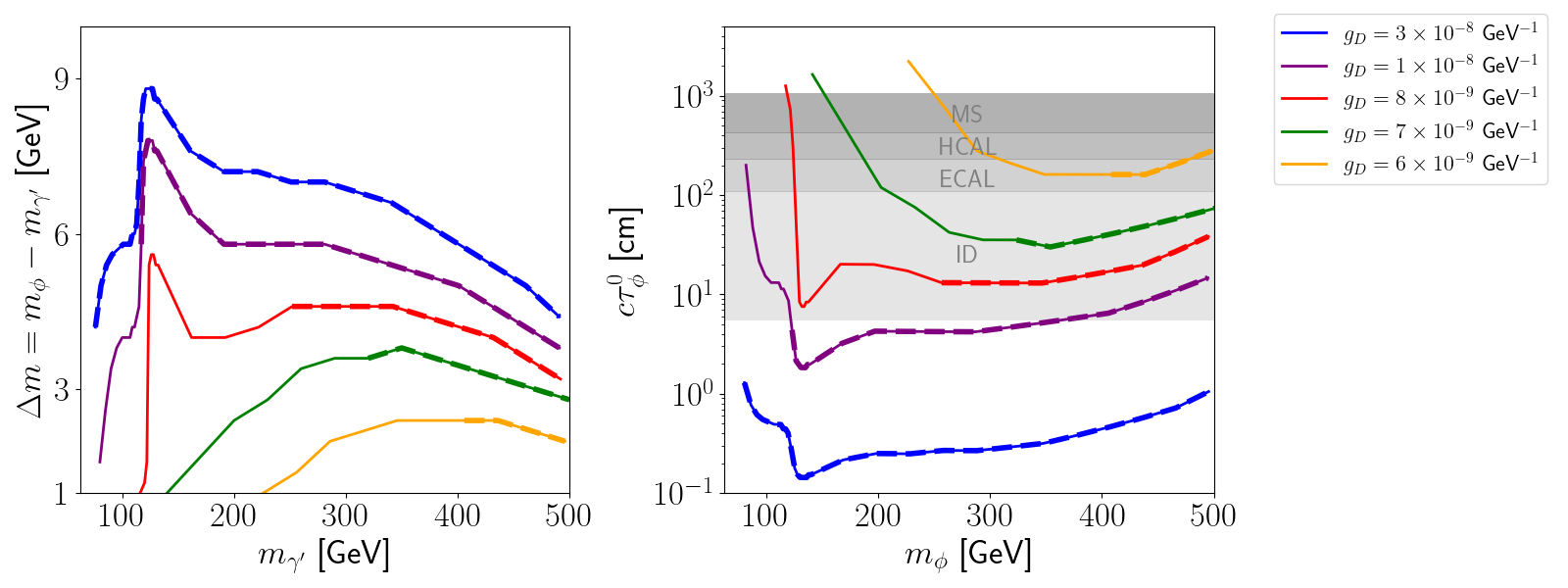}
\caption{(left) Mass parameter space with each curve fulfilling the correct relic abundance \cite{planck2018}. In the dashed part of each curve the \darkO\ function gives as maximum 50\% of the relic abundance. Here we have set $\lambda_{HS} = 1$. (right) Projection of the same curves of the plot in the left in the proper length of the ALP. The gray region corresponds to the typical distances measured by the ID, ECAL, HCAL and muon spectrometer at ATLAS at $\eta = 0$. Both plots starts the $x$-axis from $m_h/2$.} 
\label{plot5}
\end{figure}

In Fig.~\ref{plot5}, we show the results of the scan. In the plot on the left, we project the results in the $(m_{\gamma'}, \Delta m)$ plane. As shown, $\Delta m$ does not exceed $\sim$ 9 GeV, Notice that for $m_{\gamma'}\lesssim$ 120 GeV, the required $\Delta m$ tend to decrease rapidly in order to increase the cross section, since in the case of mediator FO, the channel annihilation into gauge bosons reaches its threshold, whereas in the case of coscattering, the Higgs/$Z$/$W^\pm$ bosons in the initial/final state become Boltzmann suppressed. Furthermore, the dashed part in each curve corresponds to the fact that the relic abundance given by \darkO\ is at most 50\%. In this way, the solid non-dashed lines demark the coscattering regime. 

In the plot on the right of Fig.~\ref{plot5}, we have projected the curves in the $(m_\phi , c\tau_\phi^0)$ plane. As shown, depending on the value of $g_D$, the pair of ALPs can decay in the inner detector (ID), electromagnetic calorimeter (ECAL), hadronic calorimeter (HCAL), or in the muon spectrometer (MS). A single part of the detector can test both coscattering and mediator FO at the same time, as shown by the green, red, and purple lines. The orange case (the smaller coupling) has longer lifetimes, such that the scalar may decay within the ECAL throughout the leading decay $\phi\rightarrow \gamma' + \gamma$. If $g_D$ is sufficiently large, it will cause $\phi$ to decay prompt, as shown by the blue curve. Notice that here boost factors are not taken into account, and once they considered, all the curves should lift to some extent depending on the energy of the produced pair of ALPs.

\subsection{Signals}
The signals expected by the decay of a single $\phi$ of the produced pair are given by
\begin{enumerate}
    \item Non-pointing photon plus MET.
    \item Displaced jets + MET.
    \item Displaced leptons + MET.
\end{enumerate}
Here, MET is due to the nondetection of $\gamma'$. In the first case, this type of search has been conducted at the LHC throughout the Higgs decay to LLP, which in turn decay into photons plus MET \cite{CMS:2012bbi, ATLAS:2013etx, ATLAS:2014kbb,CMS:2019zxa, ATLAS:2022vhr, ATLAS:2023meo}\footnote{Similar signals have been studied in scenarios involving heavy neutral leptons via five-dimensional operators \cite{Delgado:2022fea, Duarte:2023tdw}.}. These analyses do not apply to the parameter space studied in this work, since the Higgs pair is forbidden to decay into a pair of $\phi$ by simply kinematics. For instance, even when the search for nonpointing and delayed photons in the diphoton and missing transverse
momentum present the same signal than our scenario, the typical required cuts on MET on the outgoing two photons are above 30 and 80 GeV, respectively \cite{ATLAS:2013etx} (also see \cite{CMS:2019zxa}), too big in comparison to the typical $\Delta m$ obtained in this work (see Fig.~\ref{plot5}). Another example is the case of gauge-mediated SUSY searches \cite{ATLAS:2022zwa}, but again, the outgoing photons in our scenario are too soft to be directly recast.

The second and third types of signal are based on the three-body decay of $\phi$ through an off-shell $Z/\gamma$, with the latter producing a pair of charged SM fermions. These decay channels are suppressed compared to two-body decay due to off-shell propagators with branching ratios lower than 1\%. Even when in principle recasts are possible to constrain the parameter space of the present scenario, it presents some challenges. For instance, searches of hadronic jets plus missing energy at the muon spectrometer have been carried out \cite{ATLAS:2022gbw}, but the recast to our scenario requires a more delicate treatment, since the mediator $\Phi$ which in turn decays into long-lived scalars $s$ is always produced on-shell, unlike our case where the mediator (the Higgs boson) must be off-shell. Furthermore, the signal topology considered in \cite{ATLAS:2022gbw} does not present MET, unlike in the present case, where the DM is also present.

\begin{figure}[t!]
\centering
\includegraphics[width=0.45\textwidth]{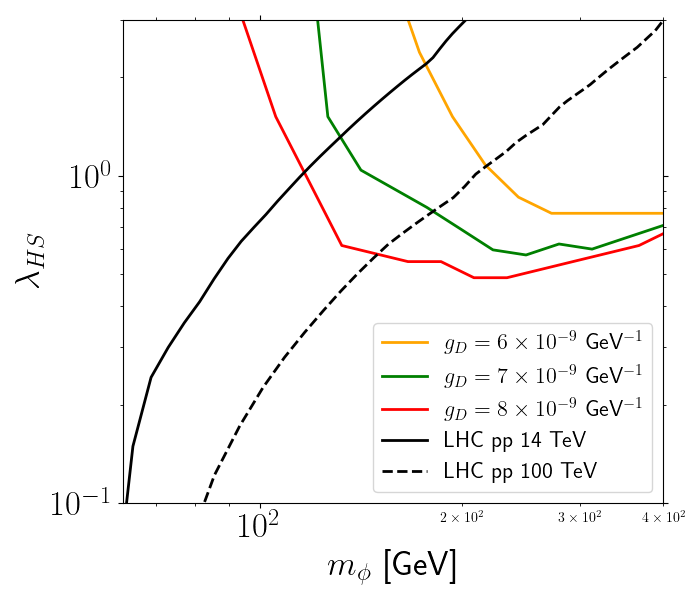}%
\caption{Parameter space for a long-lived ALP in the coscattering region. All the colored curves fulfill the correct relic abundance for $\Delta m = 1$ GeV, and the proper length of the ALPs is bigger than the reach of the ECAL, then in each case the pair of produced scalars escape of the detector as missing energy. Above each black line is excluded \cite{Craig:2014lda}.} 
\label{plot6}
\end{figure}

Instead, we use the prospects obtained in \cite{Craig:2014lda} to constrain the parameter space of our model for $m_\phi > m_h/2$ (see \cite{Baglio:2015wcg, Arcadi:2019lka} for DM searches analysis at the LHC via the Higgs portal). The simulation carried out in \cite{Craig:2014lda} considered $pp$ collisions at 14 and 100 TeV with integrated luminosity of $\mathcal{L} = 3$ and 30 ab$^{-1}$, respectively, using \mad \cite{Alwall:2011uj}. A pair of $\phi$ scalars are produced in vector-boson fusion, gluon fusion, and Higgs strahlung, with the ALPs escaping detection as missing energy. Assuming the non-observation of a significant signal, that is $S/\sqrt{S+B}$ = 1.96, neglecting systematic errors, upper bounds were constructed. In Fig.~\ref{plot6}, we show a few predictions of the coscattering regime that satisfies the correct abundance of relics (colored lines) for $\Delta m$ = 1 GeV, with the black curves the projections obtained in \cite{Craig:2014lda}. According to this, we notice that the upper bounds of the simulation exclude relevant parameter space in the low-mass region of the model, especially by the 100 TeV projection, with notorious exclusions in the plane $(m_\phi, \lambda_{HS})$ for small changes of $g_D$.
%Also see Fig.3 of 1912.08472. It is not clear to me how to extract limits on the coupling from that figure. Hay mas explicación de esto mas arribita en aquel paper.
\subsection{Prospects at MATHUSLA}
Finally, let us comment on the possibility of testing this scenario in the future MATHUSLA experiment \cite{Curtin:2018mvb, Curtin:2023skh}. This experiment is expected to be at $\sim 200$ m away from the ATLAS proton-proton interaction point. As suggested by the results of Fig.~\ref{plot5}~(right), $c\tau_0$ is not sufficient to reach $\sim 200$ m from the collision of the vertex for $\Delta m \gtrsim 1$ GeV. However, if the pair of ALPs is energetic enough, a boost factor of $\beta\gamma \approx 10$ could be enough to reach MATHUSLA, and then decay within the future detector. Furthermore, even when the decay $\phi \rightarrow \gamma + \gamma'$ will not produce any significant effect in MATHUSLA (since the experiment is sensitive to the charged particles produced), the three-body decay of $\phi$ through an off-shell $Z$ boson could introduce charged SM fermions via the chain $\phi \rightarrow \gamma' + f + \bar{f}$. The off-shellness of the $Z$ boson reduces the branching fraction of this type of channel to near 1\%. A realistic calculation of this scenario for this future experiment will be done elsewhere.

\section{Conclusions}\label{sec4}
In this work, we have investigated the thermal DM regime of a dark photon, axion-like particle, and $\gamma/Z$ boson, aka dark-axion portal, with the three particles interacting via a five-dimensional operator. We focused on the case where the dark photon plays the role of a thermal DM candidate, the ALP a slightly heavier mediator, and these dark sector particles interact with the SM via both a five-dimensional operator and the Higgs portal.

We have shown that the dark-axion portal context in the thermal regime predicts the correct relic abundance for DP masses around the electroweak scale up to the TeV scale. We have distinguished the three thermal regimes: coscattering, mediator FO, and DM (co)annihilations. We have explicitly shown the deviations of the relic abundance predictions in the coscattering regime between \darkO\ and \darkON\ functions of \micro, with the latter being the correct one, since at this level of small couplings the chemical equilibrium must be broken within the dark sector. 

At the phenomenological level, in coscattering and mediator FO regimes, the lifetime of the ALP increases enough to be considered as a LLP, presenting possible displaced vertices at the LHC detectors. We have used only simulated projections to constrain the parameter space of the model. We expect that the rich phenomenology of the dark-axion portal in the thermal DM regime motivates experimentalists to develop new strategies to search for (pseudo)scalars with masses greater than half of the mass of the Higgs boson of 125 GeV, a parameter space region barely explored. To close this paragraph, we have also shown that if the produced ALPs have a Lorentz boost factor of $\sim\mathcal{O}(10)$, they could eventually reach future detectors such as MATHUSLA.

Before closing, it is worth to emphasize about one line of research on the thermal regime of the dark-axion portal that was not worked out here. As we have discussed in the beginning of Sect.~\ref{collider_pheno}, it is possible to obtain the correct relic abundance in the sub-GeV mass range as long as $\Delta m \ll 1$ GeV and $g_D \lesssim \mathcal{O}(10^{-1})$ GeV$^{-1}$. This scenario could be interesting to explore from the point of view of direct detection (e.g. inelastic dark matter \cite{Tucker-Smith:2001myb} via SM mediators), with this type of signal eventually being tested at beam dump experiments (e.g., see \cite{Izaguirre:2015zva}).

\acknowledgments
 We would like to thank Giovanna Cottin, Paola Arias, Joel Jones Perez, Lucia Duarte, Francisca Garay, and Felix Kahlhöfer for fruitful discussions. BDS was funded by ANID (ex CONICYT) Grant No.~3220566.

%\appendix

% The bibliography will probably be heavily edited during typesetting.
% We'll parse it and, using the arxiv number or the journal data, will
% query inspire, trying to verify the data (this will probalby spot
% eventual typos) and retrive the document DOI and eventual errata.
% We however suggest to always provide author, title and journal data:
% in short all the informations that clearly identify a document.

% Please avoid comments such as "For a review'', "For some examples",
% "and references therein" or move them in the text. In general,
% please leave only references in the bibliography and move all
% accessory text in footnotes.

% Also, please have only one work for each \bibitem.
\bibliographystyle{JHEP}
\bibliography{bibliography}

\end{document}